# Scaling of the Coulomb Energy due to Quantum Fluctuations in the Charge on a Quantum Dot

L. W. Molenkamp[a,b], Karsten Flensberg[c,d], and M. Kemerink[a,*]

(a) *Philips Research Laboratories, 5656 AA Eindhoven, The Netherlands*
(b) *2. Physikalisches Institut, RWTH Aachen, D-52056 Aachen, Germany*[†]
(c) *Mikroelektronik Centret, Danmarks Tekniske Universitet, DK-2800 Lyngby, Denmark*
(d) *Danish Institute of Fundamental Metrology, DK-2800 Lyngby, Denmark*[‡]

(March 27, 1995)

The charging energy of a quantum dot is measured through the effect of its potential on the conductance of a second dot. This technique allows a measurement of the scaling of the dot's charging energy with the conductance of the tunnel barriers leading to the dot. We find that the charging energy scales quadratically with the reflection probability of the barriers. In a second experiment we study the transition from a single to a double-dot which exhibits a scaling behavior linear in the reflection probability. The observed power-laws agree with a recent theory.

Charge fluctuations reduce the effects of the Coulomb blockade [1] on electron transport in nanostructures. For example, Coulomb oscillations in the conductance of a quantum dot — periodic oscillations as a function of the voltage on an external gate electrode [1,2] —, can only be observed when the conductance $G$ of the barriers connecting the dot to external leads is reduces below $2e^2/h$, i.e. in the tunneling regime [3,4].

The influence of quantum fluctuations in the Coulomb blockade regime has recently been studied theoretically by applications of scaling- and Luttinger-liquid theory [5–9]. One of these studies [7,8] explicitly considers the case of a split-gate defined semiconductor quantum dot, and predicts a scaling of the charging energy associated with the addition of an electron to the dot with the conductance of the point-contact barriers, according to:

$$U^* = U(1-T)^{N_c}. \qquad (1)$$

Here, $U \equiv e^2/C$ is the bare charging energy (C is the self-capacitance of the dot), $U^*$ is the effective (or "renormalized") charging energy observed for finite barrier conductance, $N_c$ the total number of quantum point contacts leading to the dot, and $T$ the transmission probability of each contact. The result (1) was derived by utilizing a mapping of the two-dimensional dot-geometry to a one-dimensional model [10] with interactions equivalent to Refs. [6]. Observation of such a scaling behavior would thus be a strong support for the applicability of Luttinger-liquid theory in describing the influence of quantum fluctuations on charge transport in nanostructures.

Unfortunately, conductance measurements [3,4] are not well suited to test Eq. (1), because of the occurrence of complicating co-tunneling processes [11,12] when the barrier conductance approaches $2e^2/h$. In this Letter, we report results obtained with a fully-adjustable double quantum-dot structure, defined electrostatically in a (Al,Ga)As modulation doped heterostructure. This setup is designed to allow for a direct measurement of the charging energy $U$ as a function of barrier transparency, and does not suffer from ambiguities in the determination of $U$ due to co-tunneling processes. We discuss two experiments that probe the role of quantum charge-fluctuations in Coulomb-regulated transport. First, we use the device in an *electrometer* [13] configuration that allows a direct determination of $U$ as a function of barrier conductance. Second, we study the transition from a *single* to a *double* quantum dot geometry. We model our observations using a formalism that takes the scaling relation of Eq. (1) into account and we find good agreement between theory and experiment.

A schematic lay-out of the device is shown in Fig. 1a. The hatched areas are the TiAu gates, crosses denote ohmic contacts. The device consists of two adjacent quantum dots, 1 and 2. Gates A through F are used to define the barriers leading to the dots, and the electro-chemical potential of the quantum dots can be adjusted through gates I and II. The lithographic diameter of each dot is about 1 $\mu$m. These relatively large dimensions are necessary to minimize cross-talk between the gates; they also imply [12] that confinement effects on the transport properties are negligible, a prerequisite for the theory of Refs. [7,8]. During the experiments the samples are kept at 40 mK in a dilution refrigerator (we estimate the electron gas temperature to be 150 mK); the conductance $G \equiv G_{14}$ is measured between ohmic contacts 1 and 4, using standard low-frequency lock-in techniques.

The mode-of-operation of the electrometer experiment is schematically depicted in Fig. 1b. What is measured is the dependence of $G$ on $V_{gII}$, the voltage on gate II. In our double-dot device, scanning gate II induces Coulomb oscillations in both dot 1 and dot 2, but with a much shorter period (in $V_{gII}$) in dot 2, because of the larger dot-to-gate capacitance. The top panel of Fig. 1b shows the stepwise increase of $n_2$, the number of electrons on dot 2, within a $V_{gII}$-voltage range where $n_1$ only changes by 1. The stepwise increase of $n_2$ causes sawtooth shaped oscillations in the energy $\Delta E$ needed to add an extra electron to dot 1. As shown in the bottom panel of Fig. 1b, these oscillations are reflected in the conductance of dot 1: an additional sawtooth behavior is superimposed on the (dashed) lineshape one expects for the dot 1 Coulomb-blockade conductance peak without the capacitive coupling with dot 2. The width of the sawteeth, which can be quantified by the "back-lash" $\Delta V_{gII}$ indicated in Fig. 1b, is a direct measure of the charging energy of dot 2. Dot 1 thus acts as an electrometer [13] that measures the changes in the potential of dot 2. The full curve in Fig. 1b is, in fact, a fit to an experimental trace (dotted curve), using a theory we will discuss below. For now, we emphasize that $\Delta V_{gII}$ is proportional to $U_2$, the charging energy of dot 2.

We will now discuss an experiment that uses this proportionality to measure the scaling behavior of $U_2$ with the conductance of barriers BC and DE. In the left panel of Fig. 2 we plot $G$ versus $V_{gII}$ in a series of measurements where the barriers between dot 2 and the wide 2DEG are gradually adjusted from the metallic to the tunneling regime. From top to bottom we have $G_{BC}, G_{DE} \approx 1.3$, 0.65, 0.43, 0.14 and 0.05 $\times e^2/h$, respectively, while in all traces $G_{AB}, G_{BE}, G_{EF} \approx 0.05 \times e^2/h$, so that dot 1 is always fully in the Coulomb blockade regime. One clearly observes the sawtooth structure on the dot-1 Coulomb oscillation due to the electrometer effect. In addition, one finds that for increasing conductances $G_{BC}$ and $G_{DE}$ the sawtooth feature is much less pronounced, and, thus, that $\Delta V_{gII}$ decreases. In view of the arguments given above, it is obvious to attribute these observations to the scaling of the charging energy $U_2^*$ as a function of the conductance of barriers BC and DE. Note that the period of the sawtooth feature is unaffected by the changes in $G_{BC}$ and $G_{DE}$. We will explain below that the sawtooth-*period* is determined solely by the classical electrostatics while its *depth* (and thus $\Delta V_{gII}$) is a measure of thermal and quantum fluctuations.

In Fig. 3 we plot the values for $U_2^*/U_2$, obtained simply by measuring $\Delta V_{gII}$ in the different curves of Fig. 2, versus $(1-T)^2$, where $T$ is the conductance of barriers BC and DE, measured in units of $2e^2/h$. [$U_2$ is $\Delta V_{gII}$ measured for the bottom trace.] The excellent fit obtained is strong evidence that Eq. (1) is correct — note that $N_c = 2$ indeed corresponds to the total number of quantum points contacts connecting to the dot.

Our double-dot layout allows for a second experiment to probe the influence of fluctuations in occupation number: by adjusting the barrier BE, that connects the two dots, we can study the transition from a two-dot sample to a single-dot device. The data presented in the left panel of Fig. 4 are plots of $G (\equiv G_{14})$ versus $V_{gII}$ for various adjustments of the conductance $G_{BE}$. Barriers AB, BC, DE, and EF are all adjusted well into the tunneling regime, $G_{AB}, G_{BC}, G_{DE}, G_{EF} \approx 0.05 \times e^2/h$.

From top to bottom, $G_{BE}$ varies from 0.9 to $0.14 \times e^2/h$. Qualitatively one may interpret the data in this figure in a straightforward manner: for $G_{BE} \approx 2e^2/h$ the device may be regarded as a single dot. One then expects the rather regular, periodically spaced Coulomb oscillations [2] that are indeed observed in the top trace. On decreasing $G_{BE}$, the electrons are confined more and more to either dot 1 or dot 2. As a result, a beating pattern evolves due to the differences in capacitance between gate II and dots 1 and 2. Finally, for $G_{BE} \approx 0.01 \times e^2/h$ the device is in the limit where dots 1 and 2 are two separate (but capacitatively coupled) quantum dots, and we have returned to the situation of the electrometer experiment of Fig. 2. A quantitative discussion of the transition from one to dots must allow for fluctuations in the charge difference $(n_1 - n_2)$ between the two dots.

We have developed a theoretical model of transport through a quantum dot coupled to a second dot which includes the effects of charge fluctuations. Using this model we have produced the calculated traces in the right-hand panels of Figs. 2 and 4. We start by considering the electrostatic energy of the coupled dot system [14]:

$$E(n_1, n_2) = U_1 n_1^2 + U_2 n_2^2 + U_{12} n_1 n_2 \\ + e \sum_{i=1,2} \sum_{j=I,II} n_i a_{ij} V_{g,j} \quad (2)$$

where $n_i$ is the number of electrons on dot $i$. The constants $U_i$, $U_{12}$, and $a_{ij}$ can be expressed in terms of the elements of the capacitance matrix of the system. Let us define as $n_{i0}$ the (generally non-integer) number of electrons on dot $i$ that minimizes $E(n_1, n_2)$. We now may write the dependence of $E(n_1, n_2)$ on small deviations $\delta n_i \equiv n_i - n_{i0}$ as

$$E(n_1, n_2) = E(n_{10}, n_{20}) + \delta E(n_1, n_2), \quad (3a)$$
$$\delta E(n_1, n_2) = U_1 \delta n_1^2 + U_2 \delta n_2^2 + U_{12} \delta n_1 \delta n_2. \quad (3b)$$

Here $\delta E(n_1, n_2)$ is the quadratic term which controls the fluctuation away from the optimum charge configuration $(n_{10}, n_{20})$. All fluctuation-dependent properties (such as the transport properties) are thus periodic functions of $n_{10}$ and $n_{20}$. Consequently, the periodicity of the Coulomb oscillations is *unaffected* by number fluctuations. [Note that this formalism treats occupation-number fluctuations of quantum-mechanical and thermal origin on an equal footing.] An increase in number fluctuations due to lowering of the tunnel barriers may be thought of as a decrease of the charging energies that enter $\delta E(n_1, n_2)$. With this in mind, we model the effect of quantum fluctuations on the transport properties by invoking the renormalized charging energy of Eq. (1), as follows. We replace the deviation $\delta E(n_1, n_2)$ in Eq. (3) by its "renormalized" counterpart $\delta E^*(n_1, n_2)$, given by a similar expression as in Eq. (3b), but with the relevant $U_i$ replaced by renormalized charging energies $U_i^*$ (and leaving $n_{i0}$ unrenormalized). Within this model we can generalize the rate equation approach of Refs. [15] to our double dot system and then solve for the linear conductance. We obtain

$$G = \tfrac{1}{2} G_0 \beta \sum_{n_1,n_2} W_0^*(n_1, n_2) \times$$
$$f[\delta E^*(n_1, n_2) - \delta E^*(n_1 - 1, n_2)], \quad (4a)$$

$$W_0^*(n_1, n_2) = \frac{\exp[-\beta \delta E^*(n_1, n_2)]}{\sum_{n_1,n_2} \exp[-\beta \delta E^*(n_1, n_2)]}, \quad (4b)$$

$$f(E) = E/(1 - e^{-\beta E}), \quad (4c)$$

where $\beta = 1/k_B T$ and $G_0 = G_{AB} = G_{EF}$. We now proceed to find expressions for $\delta E^*(n_1, n_2)$ that are applicable to our experiments.

*First experiment.* — The case of the electrometer measurement of Fig. 2 is straightforward: when barriers BC and DE are opened, the charge on dot 2 is allowed to fluctuate more and more. This means that $U_2$ is renormalized, and Eq. (3b) becomes

$$\delta E^*(n_1, n_2) = U_1 \delta n_1^2 + U_2^* \delta n_2^2 + U_{12} \delta n_1 \delta n_2 \quad (5)$$

The right panel of Fig. 2 shows lineshapes calculated from Eqs. (2), (4) and (5). In order to obtain a consistent set of fits, we first determine the parameters $U_i = 0.13$ meV, $U_{12} = 0.009$ meV, $a_{ii} = -0.20$, and $a_{12} = -3.12$ from a fit of Eqs. (2) and (4) to the bottom trace of the left panel of Fig. 2 — which is the same as the experimental (dotted) trace in the bottom panel of Fig. 1b —, where both dots are fully in the tunneling regime. The upper curves are then obtained from Eqs. (4) and (5), keeping the same values for $U_1$, $U_{12}$, and $a_{ij}$ while varying $U_2^*$ with $G_{BC}, G_{DE}$ according to Eq. (1). As is evident from the theoretical curves, this procedure yields a very good agreement with the experiments.

*Second experiment.* — The gradual transition between single- and double-dot behavior of $G$ in Fig. 4 can also be modeled within this approach: When barrier BE is opened the charge difference between the two dots, $n_1 - n_2$ can fluctuate. This again leads to an expression for $\delta E^*(n_1, n_2)$:

$$\delta E^*(n_1, n_2) = U_+(\delta n_1 + \delta n_2)^2 + U_-^*(\delta n_1 - \delta n_2)^2$$
$$+ \Delta U(\delta n_1 + \delta n_2)(\delta n_1 - \delta n_2), \quad (6)$$

where we have defined $U_+ = (U_1 + U_2 + U_{12})/4$, $U_- = (U_1 + U_2 - U_{12})/4$, and $\Delta U = (U_1 - U_2)/2$. Note that only for $U_-^*$ the renormalized value is used.

The curves in the right panel of Fig. 4 are calculated from Eqs. (4) and (6), using Eq. (1) to describe the scaling behavior of $U_-^*$. Note that $N_c = 1$ in this case because only one point contact is opened. Once again, we have obtained a very reasonable agreement with the experiment, which we regard as an extra indication of the validity of Eq. (1). However, we also note that our calculations display an increase in the width of the dot 1 Coulomb resonance with increasing $G_{BE}$, which is not observed experimentally. The reason for this discrepancy is presently unclear.

In conclusion, we have performed and analyzed experiments aimed at understanding the role of charge fluctuations in the transport properties of quantum dots. We find that the dependence of the charging energy of a quantum dot on the conductance of the point contact tunnel barriers can be well described using a scaling equation. It would be useful to verify the validity of the scaling equation for other power laws, which could be accomplished, e.g., by performing experiments in a high magnetic field, or by varying only one of the tunnelbarriers.

We thank B. W. Alphenaar, B.Y.-K. Hu, M. J. M. de Jong, A. A. M. Staring and N. C. van der Vaart for stimulating discussions, and O. J. A. Buyk and M. A. A. Mabesoone for technical assistance. The heterostructures were grown by C. T. Foxon at Philips Research Laboratories in Redhill (Surrey, UK).

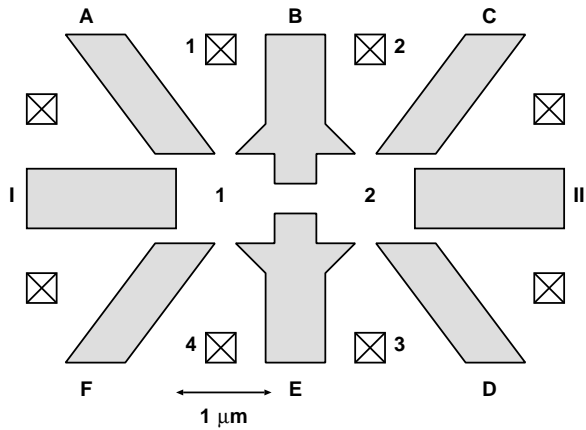

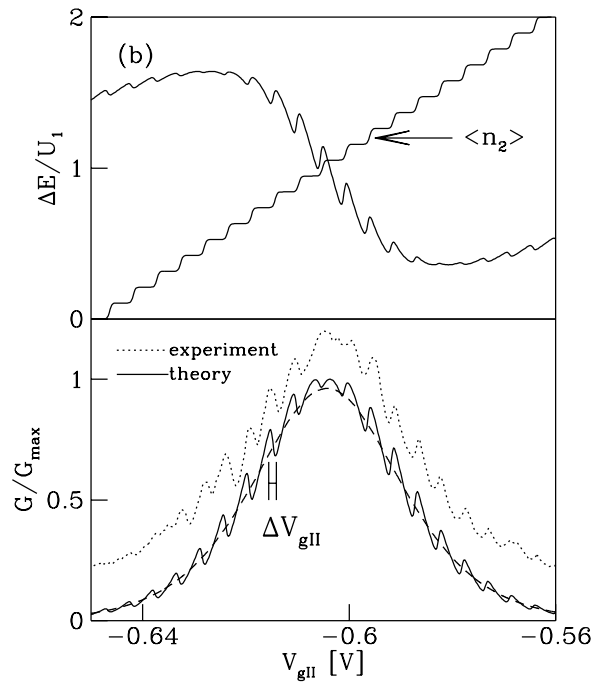

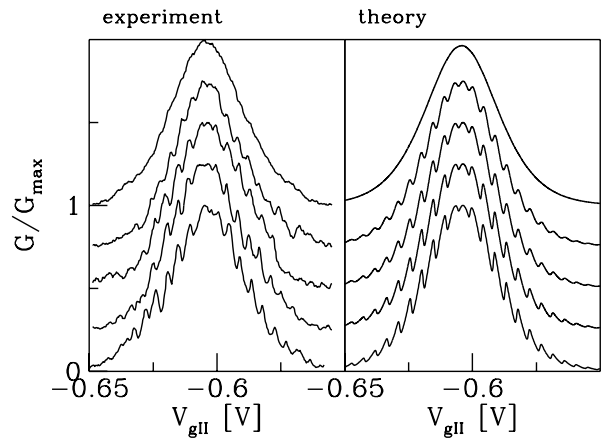

FIG. 1. (a) Schematic lay-out of the double dot sample. The hatched areas denote gates, the crosses ohmic contacts. (b) The operating principle of the electrometer experiment: scanning $V_{gII}$ leads to an increase in $n_2$, the number of electrons on dot 2. The concomitant sawtooth oscillations in $\Delta E$, the energy needed to add an extra electron to dot 1, resulting in a back-lash $\Delta V_{gII}$ in the Coulomb oscillation of dot 1. Full lines are results of the model calculation and the bottom panel shows a fit to the experimental trace (dotted curve). The dashed line corresponds to the case with where there is no charging energy for the second dot. $\Delta V_{gII}$ thus measures the effective charging energy of dot 2, which is used for determining $U_2^*$, as plotted in Fig. 3.

FIG. 2. Traces of $G$ versus $V_{gII}$ in an electrometer experiment, where the conductance of barriers BC and DE is varied. An offset of $0.25 \times G/G_{max}$ is used between consecutive curves. Left panel: Experimental data, where from top to bottom $G_{BC}, G_{DE} \approx 1.3$, 0.65, 0.43, 0.14 and $0.05 \times e^2/h$. In all traces $G_{AB}, G_{BE}, G_{EF} \approx 0.05 \times e^2/h$. Right panel: The results of calculations using the renormalized $\delta E^*$ of Eq. (5).

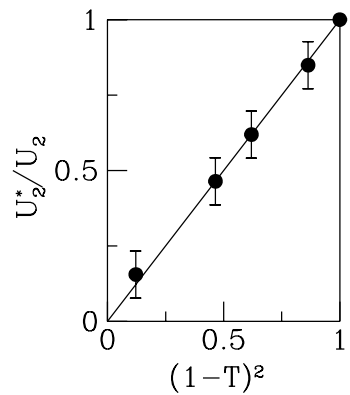

FIG. 3. Plot of the ratio $U_2^*/U_2$ between "renormalized" and bare charging energy of dot 2, versus $(1-T)^2$, where $T$ is the transmission of barriers BC and DE that control the coupling between dot 2 and the external leads. The values for $U_2^*/U_2$ are determined from the $\Delta V_{gII}$ as shown in Fig. 1b. The linear dependence found in this plot indicates the validity of a scaling law of the type of Eq. (1).

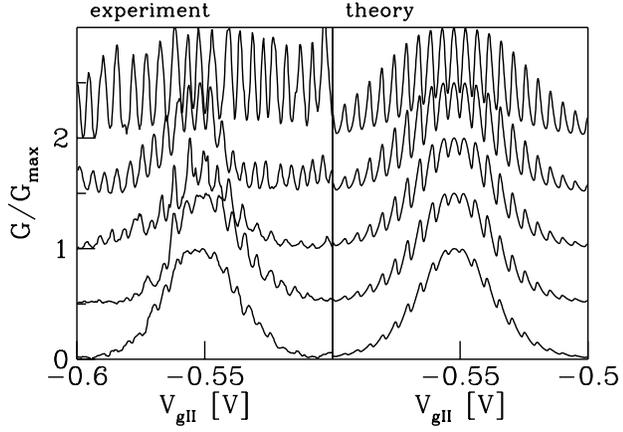

FIG. 4. $G$ versus $V_{gII}$ in an experiment where $G_{BE}$ is varied, so that the device changes from a single to a double quantum dot. An offset of 0.5 $\times G/G_{max}$ is used between consecutive curves. Left panel: Experimental traces, where, from top to bottom, $G_{BE} = 0.9$, 0.5, 0.26, and 0.14 $e^2/h$, respectively. During this experiment all other tunnelbarriers are kept at a conductance of about 0.05 $e^2/h$. Right panel: Theoretical curves using Eq. (6) for $\delta E^*$.